# PREVENTION OF TERRORIST CRIMES IN THE NORTH CAUCASUS REGION

**Ivan Kucherkov,**
*Associate Professor, Russian University of Transport,*
*Moscow, Russia*

**Mattia Masolletti,**
*Associate Professor, NUST University*
*Rome, Italy*

**Abstract**

The relevance of the topic is dictated by the fact that in recent decades, the threat to international security emanating from terrorism has increased many times. Terrorist organizations have become full-fledged subjects of politics on a par with political parties. In addition, enormous power and resources are concentrated in the hands of terrorist groups. Terrorist activity has become the usual way of leading a political struggle, expressing social protest. In addition, terrorism has become a tool in economic competition. Each terrorist action entails more and more human casualties. It breeds instability, fear, hatred, and distrust in society. The authors pay special attention to counter-terrorism activities in the North Caucasus Region.

1. **Introduction**

Currently, the actions of extremist and terrorist organizations constitute one of the most likely and actual threats to security in the modern world. For instance, in the National Security Strategy of the Russian Federation till 2020, terrorist activities are recognized as one of the most serious threats to the national security of a country. In the second part of this document, which is devoted to the security of the state and society, Article 37 lists the main sources of threats to Russia. 'The main sources of threats to national security in the sphere of state and public security are: ... activities of terrorist organizations, groups and individuals aimed at forcibly changing the foundations of the constitutional system of the Russian Federation, disrupting the normal functioning of state bodies (including violent actions against state, political and public figures), the destruction of military and industrial facilities, enterprises and institutions, ensuring the functioning of society, intimidation of the population, including through the use of nuclear and chemical weapons or dangerous radioactive, chemical and biological substances ....' [1].

2.   **Main part**

Terrorism today has become an integral companion of our lives, causing worldwide fear, insecurity in the future and a huge political and economic damage. However, the phenomenon of world scale and the global threat of modernity, terrorism can be called only from the beginning of the 90s XX century, when he went beyond the local phenomena and became international. Let us turn to statistical data: while in 1980, 500 were recorded in all countries of the world, by 2006 there were 14,338 acts of terrorism. As we see, terrorism is now time is transnational.

| Country | Overall Score | Regional Rank | Change 2002-2018 | Change 2016-2018 |
|---|---|---|---|---|
| Ukraine | 5.547 | 24 | 3.961 | -0.501 |
| Russia | 4.900 | 37 | -1.933 | -0.330 |
| Tajikistan | 3.947 | 50 | 1.212 | 1.714 |
| Kazakhstan | 1.566 | 85 | 1.184 | -0.662 |
| Kyrgyz Republic | 1.467 | 87 | -0.340 | -0.252 |
| Georgia | 1.335 | 90 | -1.498 | -0.087 |
| Armenia | 1.173 | 94 | 0.053 | -0.519 |
| Azerbaijan | 0.698 | 103 | -0.868 | -0.259 |
| Moldova | 0.115 | 123 | 0.077 | -0.114 |
| Uzbekistan | 0.019 | 135 | -2.068 | -0.019 |
| Belarus | 0.000 | 138 | -0.229 | 0.000 |
| Turkmenistan | 0.000 | 138 | -0.229 | 0.000 |
| Regional average | | | -0.057 | -0.086 |

Fig. 1. Russia and Eurasia Global Terrorism Index score, rank and change in the period of 2002-2018
*Source: Global Terrorism Index, 2002-2018.*

Many scientifically based predictions about the development of mankind in the next few decades are already being made, and many scientists point out that the main sources of international terrorism will be countries with insufficiently strong government, ethnic, cultural or religious friction, weak economies and poorly guarded borders. At the same time, the threat of using new high-speed transmission of information and other technological advances to unite the illegal activities of transnational terrorist networks, and the assumption made by Russian experts Ursul and Romanovich that international terrorism is capable of provoking a war of civilizations with its disastrous consequences is quite reasonable, especially considering the modern realities of terrorism [2].

Terrorism is a programmatically implemented sequence of violent destructive actions to achieve certain political and other goals, aimed at destabilizing and intimidating society [8-13].

There are various definitions of this phenomenon, but they all prove that terrorism, first of all, is intimidation, instilling fear, creating an atmosphere of horror to achieve any goals, in particular, making certain decisions by a third party, although. In some cases, the requirements are not formulated and even a demonstration of oneself, one's strength and capabilities can take place, and thus deterrence is achieved.

The act of terrorism of an international character could be understood as:

1. violent acts in the form of an attempt or commission of an attack, seizure, abduction, bodily harm, murder or actions that pose a threat to officials and their families in the field of political, economic, technical, commercial and cultural relations between subjects of international law;
2. the seizure, damage and destruction of property necessary for the implementation of political, economic, technical, commercial and cultural relations between states, as well as means, equipment and facilities for air, water, rail and road transport, if these actions were characterized by the presence of an international element.

Professor Zhdanov emphasizes that an international element means the commission of a terrorist act [3]:

1. on the territory of one or several states or on a territory not falling under the jurisdiction of any state;
2. a foreign citizen or nationals or the complicity of a foreign citizen or national takes place;
3. in relation to a foreign citizen or property of a foreign natural, legal person or state.

Based on the above material, it can be concluded that international terrorism is the organization and implementation of deliberate, unlawful violent acts (actions) or the threat of their use, carried out with the aim of violating international security, intimidating the population or influencing government decisions that satisfy interests of terrorists.

The main object of international terrorist acts is public (international) security.

According to Zhdanov [3], it is advisable to divide all the objects of terrorist actions into primary and secondary. Primary objects include individual individuals, groups of individuals, or material objects. To the secondary should include the objects of management, which can be called as social relations in general.

The main objects of international terrorism are:
1. individual individuals, groups of persons or material objects;
2. individual states, unions and associations of states;
3. international organizations;
4. the world community as a whole.

The main subjects of international terrorism are:
1. non-state terrorist structures, as a rule, extremist international movements of the left and right orientation, nationalist or religious-political sense;
2. organized criminal associations engaged in illegal foreign trade activities, such as drug trafficking, international criminal mercenaries;
3. individuals or groups of individuals acting independently or according to state instructions;
4. state officials, as well as state structures of various countries, carrying out secret or obvious terrorist actions against other states or against citizens of another state, as well as supporting terrorist groups whose activities are aimed at opposing those or other political or economic systems.

It should be noted that international terrorism occurs when [4]:
1. a terrorist and persons suffering from a terrorist act are citizens of the same state or of different states, but the crime is committed outside these states;
2. a terrorist act directed against internationally protected persons;
3. preparation for a terrorist act is conducted in one state, and carried out in another;
4. having committed a terrorist act in one state, a terrorist takes refuge in another, and the question arises of his extradition (extradition).

Thus, the goal of individual terrorist acts can be provocation of international conflicts, violent change or undermining the socio-political system of sovereign states, destabilization and overthrow of their legitimate governments, violent opposition to peoples' self-determination, establishing more favorable conditions for national and transnational corporations by eliminating economic competitors.

Terrorists act systematically: at the beginning of a terrorist action, the purpose of which is to intimidate the population, and then comes the stage of manipulating the consciousness and behavior of people to realize their selfish goals.

In a broad sense, terrorism is a combination of extremist ideology, a set of organizational structures and special practices for carrying out actions that destabilize public order.

The extremist ideology underlying international terrorism sets the following objectives:

1. drawing attention to the problems of a large social group or social stratum of other states and peoples and the world community as a whole, unsatisfied with their position;
2. to force the authorities of individual countries or a union of states to take into account the interests of terrorists;
3. in order to develop the activities of international terrorist organizations against the authorities both at the national and international level, to strengthen and expand their ranks, it becomes necessary, with the help of ideology, to win over a large number of active supporters;
4. an attempt to justify their terrorist actions by the rest of the world community.

It is necessary to understand that the globalization process has expanded the so-called working space for terrorists: obtaining open access to new transport, information and communication and other resources has ensured the commission of terrorist acts on any continent or territory.

The transformation of the problem of international terrorism into one of the most acute global problems of our time is due to the following reasons:

- first, the goals of international terrorists and transnational organized crime coincide - the elimination of the existing mechanisms for the functioning of the world system and the construction of a new world order based on violence and fear;

- secondly, classical terrorist groups, realizing that during the acceleration of the globalization processes they will not be able to quickly adapt to the new realities of international relations, they join a more organized, highly professional transnational terrorist network with its own infrastructure, logistics and sources of financing;

- thirdly, international terrorism applies radical religious and political concepts as its ideological 'feed' and method of recruiting new militants; in addition, terrorists use religious tenets to justify their actions and sacrifice the meaning of the terrorist acts committed by them;

- finally, in modern conditions the so-called border between state and international terrorism has been almost completely eliminated.

The phenomena of international terrorism and extremism are closely interrelated. Both international terrorism and extremism are characterized by a commitment to extreme views and measures. It should also be noted that terrorism itself is most often the ultimate form of extremism.

The interrelation and mutual influence of international terrorism and extremism, in our opinion, are manifested in the following:

1) The territories where extremism develops become a fruitful ground and a springboard for the development of international terrorism;

2) Both international terrorism and extremism are phenomena that are used as 'political weapons' by various countries to realize their geopolitical and other interests;

3) International terrorism feeds the forces of extremists in certain regions through financial investments and the creation of a network for the training of professional fighters;

4) Both international terrorism and extremism are based on such motivational characteristics as ethnicity and religiosity. Interethnic and interfaith contradictions help the development of both extremism and international terrorism.

Extremist organizations are not independently operating groups in isolation from the 'basic', theoretical fundamentalist organizations, but they are their 'combat wings'. It is Islamic organizations that have created their own alternative to secular civil society in the form of institutions such as 'private mosques, professional associations, trade unions, hospitals, Islamic banks and schools that provide Islamic fundamentalism with an inexhaustible source of recruits and allow it to spread its influence almost everywhere'. The central formula of fundamentalism: the cause of all evils lies in the imperfection and corruption of all secular modernists, and the only way out is in the Islamic model of society based on Sharia.

Extremism and terrorism have national roots, but they develop and survive, relying on international support. The extremists in the North Caucasus would not have been able to fight the federal center for such a long time if they had not received support from various international terrorist organizations.

The North Caucasus today is a complex system of relations between peoples and states. The Caucasus is not only a natural geographical boundary between Europe and the Asian and Middle Eastern regions, the oldest transport artery (the Small Silk Road, the Road of Spices and Incense) connecting the two continents, but also a source of strategic natural energy resources-oil and gas. And, finally, the Caucasus is the junction of two civilizations-Western (Christian) and Muslim, two world ideologies that declare different spiritual values and world order.

Since in the new geopolitical conditions, the main redistribution of the world is carried out for the establishment of control over natural resources, geostrategic and naval routes, the main strategic goal of the major world powers is to push Russia to the northeast of Eurasia, away from one of the main communication approaches to the center of world resources-the Mediterranean - Black Sea-Caucasus – Caspian region [5]. The immediate goals are to make the existing Russian main gas and oil pipelines uncompetitive and costly by bypassing Russia's own transport arteries that deliver cheap raw materials from Kazakhstan, Azerbaijan, Iran to Europe and America.

Therefore, the internal political instability in the North Caucasus and Transcaucasia is largely beneficial to a number of state and interstate entities that pursue exclusively their national (political and economic) interests.

Many countries have announced their strategic claims to this region, called the 'solar plexus of Eurasia', and, first of all, the United States, Great Britain, and Turkey.

How to prevent terrorist crimes in the North Caucasus region? First of all, the law enforcement agencies and special services of Russia need to take comprehensive measures to destroy the infrastructure for training 'suicide bombers' both in Russia and abroad, additional measures to prevent potential terrorist acts at important and high-security facilities of the military-industrial complex (MIC), 'Rosatom', transport, life support facilities, other critical facilities, places of mass stay of civilians throughout Russia [6].

Also, it is necessary to use more widely, including on distant approaches to the transport infrastructure, video systems of domestic and foreign developers with intelligent detection of images (photos) of wanted participants in organized terrorist activities and their accomplices, with the output of information to the monitors of situational crisis centers (SCC) for making immediate management decisions to neutralize terrorist threats.

It is important to prevent crimes of a terrorist nature, including terrorist acts, and first of all by increasing the effectiveness of the operational-search, intelligence and counterintelligence activities of the subjects of the ORD (the Federal Security Service of Russia, the Ministry of Internal Affairs of Russia, and etc.), improving agent work and undermining the financial foundations of organized criminal activity, including terrorist. The terrorist infrastructure in the North Caucasus has not been completely destroyed, the military organizational and staff structure, which has a hierarchical-network character, has been preserved in the activities of organized criminal formations (OCF), and its members have switched to guerrilla methods of fighting against law enforcement agencies and special services of Russia, striking at the most vulnerable places, and also partially penetrated into state structures [7].

### 3. Conclusion

Officials who carry out operational and combat prevention and suppression of terrorist financing solve these tasks through personal participation in it, by organizing and conducting operational and combat events, using the help of officials and specialists with scientific, technical and other special knowledge, as well as individual citizens with their consent on a public and secret basis.

Operational-combat prevention and suppression of terrorism can be carried out both by conducting operational-combat measures by special police units, and by operational implementation units. Special police units are suppressing the financing of terrorist groups, destroying underground oil refining plants, crops of drug-containing crops, stopping the transportation of seafood, large sums in foreign currency and rubles. When freeing hostages taken for ransom, as well as suppressing the slave trade.

The implementation of such measures for the operational and combat prevention and suppression of terrorist financing, such as the prompt detention of couriers with money, the red-handed capture of leaders and active participants of terrorist groups during settlements during criminal transactions, during an operational strike and the seizure of money intended for terrorism, makes it possible to oppose new, more daring and dangerous manifestations of terrorist financing, new forms and methods of power decisions on the part of internal affairs bodies. The model of the organization of operational and combat prevention and suppression of terrorist financing developed within the framework of this theory allows solving a number of applied tasks:

- to develop a set of system basic methods of operational and combat prevention and suppression of the financing of terrorist activities, operational search measures to detect financing channels and capture (eliminate) terrorist groups controlling the financing channel, to promptly detain (capture) their leaders and the most active members, to withdraw funds from them intended for the financing of terrorism;

- to develop proposals on the implementation of the methods of operational and combat prevention and suppression of terrorism in the practical activities of the Department of Internal Affairs;

- to prepare methodological recommendations for the operational and combat prevention and suppression of the internal affairs of terrorist financing;

- to develop proposals for a number of organizational and legal internal departmental documents regulating operational and combat prevention and suppression of terrorist financing by internal affairs bodies and the practice of training employees of special forces of internal affairs bodies to carry out this activity.

The definition of the general provisions of the organization of operational-combat prevention and suppression of terrorist financing and the essence of the organization of a single operational-combat event to suppress a specific channel of terrorist financing shows that the tactics of combating terrorism are filled with operational measures. The latter allow you to coordinate the efforts of employees, individual functions and elements of an operational and combat unit in solving a number of issues.

The Federal Law 'On Operational Search Activities' has established an exhaustive list of operational search activities. The list of operational and combat measures has been developed by the theory of operational and combat activities of the internal affairs bodies, and it includes: operational detention; special events; special actions-the capture (destruction) of leaders and other (if necessary) gang members, the destruction of warehouses with weapons and ammunition; special operations; operational environment; operational strike.

# References

[1] The National Security Strategy of the Russian Federation till 2020, article 37.

[2] Ursul A.D., Romanovich A. L. (2001) Security and sustainable development. Moscow, p. 426.

[3] Zhdanov H. B. (2003) The Islamic concept of the world order. - M. p. 173

[4] Katzman K. Terrorism: Middle Eastern Groups and State Sponsors. – L., 2002.

[5] Pinchuk A. Yu. (2018) On the problem of understanding the essence and specifics of international terrorism in the modern world // Azimut of scientific research: economics and management: political sciences. - Vol. 7. - № 2(23). - Pp. 396-399;

[6] Karyagina O. V., Bernatsky D. S. (2020) Extremism and terrorism: differences of concepts through their legislative reflection and correlation with antisocial behavior // Science and modernity: proceedings of the All-Russian scientific and Practical Conference of students and young scientists. – Taganrog. - Pp. 60-62.

[7] Erokhin D. V. (2018) International legal bases of countering terrorism // Bulletin of Omsk University. - The 'Law' series. - № 2(55). - Pp. 185-190.

[8] Laquer W. (1977) Terrorism. - Boston, MA: Little & Brown.

[9] Laquer W. (1977) Guerrilla: A Historical and Critical Study. - London: Weidenfeld & Nicolson, 1977.

[10] Jenkins B.M. (1974) International Terrorism: A new kind of warfare. – RAND. - P.4.

[11] Jenkins B.M. (1980) The Study of Terrorism: Definitional Problem. - Santa Monica (California).

[12] Bell J. B. (1975) Transnational Terror. Wash.: Hoover pol. Studies.

[13] Schmid A. (1982) Violence as Communication: Insurgent Terrorism and the Western News Media. – Sage.